\documentclass[english]{revtex4}
\usepackage[T1]{fontenc}
\usepackage[latin1]{inputenc}
\usepackage{float}
\usepackage{amsmath}
\usepackage{graphicx}

\makeatletter

\usepackage{babel}
\makeatother
\begin{document}
\begin{flushright}JLAB-THY-04-220 \\
 April 12, 2004  
\end{flushright}

\vspace{2cm}

\title{Inclusive Photoproduction of Lepton Pairs in the Parton Model%
\footnote{Poster presentation given at V Latin American Symposium on Nuclear
Physics. To be published in the Proceedings.%
}}

\author{A. PSAKER}

\bigskip

\affiliation{Theory Group, Jefferson Laboratory,\\
 Newport News, VA 23606, USA\\
 and\\
 Physics Department, Old Dominion University,\\
 Norfolk, VA 23529, USA}

\begin{abstract}
\vspace{2cm}

In the framework of the QCD parton model, we study unpolarized scattering
of high energy real photons from a proton target into lepton pairs
and a system of hadrons. For a given parametrization of parton distributions
in the proton, we calculate the cross section of this process and
show the cancellation of the interference terms.

\vspace{5mm}

PACS number(s): 13.40.-f, 13.60.Fz, 13.60.Hb
\end{abstract}
\maketitle
\newpage

\section{Introduction}

The inclusive virtual Compton scattering is the reaction in which
a high energy lepton beam bombards a proton target and scatters off
the target inelastically

\begin{eqnarray}
l+p & \rightarrow  & l'+\gamma +X,\label{eq:inclusivevcs}
\end{eqnarray}
 where $l$ and $l'$ represent either an electron or a muon and \emph{X}
a system of the final state hadrons. This process was studied in the
parton model by Brodsky et al. \cite{Brodsky:1972yx}. The basic idea
of the model is the assumption that at high energies the constituents
of hadrons behave as if they are free point-like particles. Thus,
one picks up only the lowest order electromagnetic contributions to
the cross section and neglect all QCD corrections. In the present
paper we consider inclusive photoproduction of lepton pairs

\begin{eqnarray}
\gamma \left(k\right)+p\left(P\right) & \rightarrow  & l^{-}\left(p_{1}\right)+l^{+}\left(p_{2}\right)+X,\label{eq:inclusiveleptonpairphotoproduction}
\end{eqnarray}
 which is a crossed process to Eq.(\ref{eq:inclusivevcs}). At the
level of the elementary photon-parton scattering subprocess, the incident
photon can either scatter off a parton or split into a lepton pair.
We call these contributions the Compton and the Bethe-Heitler process,
respectively. Since we imagine the reaction to occur at very high
energy, we can assume that all the relevant parton masses are negligible.

\section{The Compton Process}

\begin{figure}[H]
\begin{center}

\includegraphics[  scale=0.5]{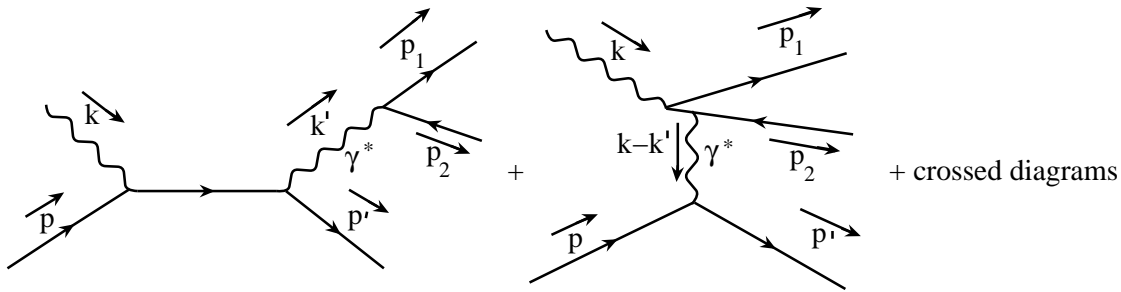}

\end{center}\caption{Feynman diagrams for Compton and Bethe-Heitler process. In two crossed diagrams the real and virtual photons are interchanged.}

\label{diagrams}
\end{figure}

There are two Feynman diagrams corresponding to the Compton contribution
depicted in Fig. \ref{diagrams}. Using $\epsilon _{\nu }(k)$ and
$Q_{a}$ to denote the polarization vector of the initial photon and
the electric charge of the parton of type \emph{a} in units of |\emph{e}|
we write the invariant matrix element as

\begin{eqnarray}
M_{C}=M_{C1}+M_{C2} & = & e^{3}Q_{a}^{2}\frac{1}{k'^{2}}\overline{u}(p_{1})\gamma _{\mu }v(p_{2})\epsilon _{\nu }(k)\overline{u}(p')\left[\frac{\gamma ^{\mu }\not \! k\gamma ^{\nu }+2\gamma ^{\mu }p^{\nu }}{2\left(p\cdot k\right)}+\frac{-\gamma ^{\nu }\not \! k'\gamma ^{\mu }+2\gamma ^{\nu }p^{\mu }}{-2\left(p'\cdot k\right)}\right]u(p).\label{eq:comptonamplitude}
\end{eqnarray}
 Averaging the squared amplitude of Eq.(\ref{eq:comptonamplitude})
over the initial parton and photon polarizations and summing over
the final lepton and parton polarizations, one gets

\begin{eqnarray}
\frac{1}{4}\sum _{polarizations}|M_{C}|^{2} & = & \frac{e^{6}Q_{a}^{4}}{4k'^{4}}L_{\mu \rho }H^{\mu \rho },\label{eq:averagedcomptonamplitudesquared}
\end{eqnarray}
 where the leptonic and hadronic tensors are

\begin{eqnarray}
L_{\mu \rho }=4\left[p_{1\mu }p_{2\rho }+p_{1\rho }p_{2\mu }-g_{\mu \rho }\frac{k'^{2}}{2}\right], &  & H^{\mu \rho }=4\left\{ \frac{A_{C}}{\left(p\cdot k\right)^{2}}+\frac{B_{C}}{\left(p'\cdot k\right)^{2}}+\frac{C_{C}}{\left(p\cdot k\right)\left(p'\cdot k\right)}\right\} ,\label{eq:comptontensors}
\end{eqnarray}
 with the coefficients

\begin{eqnarray}
A_{C} & = & \left(p\cdot k\right)\left[p'^{\mu }k^{\rho }+p'^{\rho }k^{\mu }-g^{\mu \rho }\left(p'\cdot k\right)\right],\nonumber \\
B_{C} & = & \left(p\cdot k'\right)\left[p'^{\mu }p^{\rho }+p'^{\rho }p^{\mu }\right]+\left(p'\cdot k'\right)\left[p^{\mu }k'^{\rho }+p^{\rho }k'^{\mu }-2p^{\mu }p^{\rho }\right]+\left(p\cdot p'\right)\left[2p^{\mu }p^{\rho }-p^{\mu }k'^{\rho }-p^{\rho }k'^{\mu }\right]\nonumber \\
 &  & +\frac{k'^{2}}{2}\left[-p'^{\mu }p^{\rho }-p'^{\rho }p^{\mu }+g^{\mu \rho }(p\cdot p')\right]-g^{\mu \rho }\left(p\cdot k'\right)\left(p'\cdot k'\right),\nonumber \\
C_{C} & = & \left(p\cdot k\right)\left[p'^{\mu }k'^{\rho }+p'^{\rho }k'^{\mu }-p'^{\mu }p^{\rho }-p'^{\rho }p^{\mu }\right]+\left(p'\cdot k\right)\left[2p^{\mu }p^{\rho }-p^{\mu }k'^{\rho }-p^{\rho }k'^{\mu }\right]\nonumber \\
 &  & +\left(p\cdot k'\right)\left[-p'^{\mu }p^{\rho }-p'^{\rho }p^{\mu }-p'^{\mu }k^{\rho }-p'^{\rho }k^{\mu }\right]+\left(p'\cdot k'\right)\left[2p^{\mu }p^{\rho }+p^{\mu }k^{\rho }+p^{\rho }k^{\mu }\right]\nonumber \\
 &  & +\left(p\cdot p'\right)\left[-p^{\mu }k^{\rho }-p^{\rho }k^{\mu }-p^{\mu }k'^{\rho }-p^{\rho }k'^{\mu }\right]\nonumber \\
 &  & +g^{\mu \rho }\left[2\left(p\cdot p'\right)\left(p\cdot k'\right)+\left(k\cdot k'\right)\left(p\cdot p'\right)+\left(p\cdot k'\right)\left(p'\cdot k\right)-\left(p\cdot k\right)\left(p'\cdot k'\right)\right].\label{eq:comptoncoefficients}
\end{eqnarray}
 The next step is to integrate Eq.(\ref{eq:averagedcomptonamplitudesquared})
over the Lorentz-invariant phase space in a specific frame of reference,
i.e.

\begin{equation}
\frac{1}{\left(2\pi \right)^{5}}\int \left(\frac{d^{3}p'}{2E_{p'}}\right)\int \left(\frac{d^{3}p_{1}}{2E_{1}}\right)\int \left(\frac{d^{3}p_{2}}{2E_{2}}\right)\; \delta ^{(4)}\left(p+k-p'-p_{1}-p_{2}\right)\frac{e^{6}Q_{a}^{4}}{4k'^{4}}H_{\mu \rho }L^{\mu \rho }.\label{eq:invariantphasepsaceintegral}\end{equation}
 Since $L_{\mu \rho }$ depends only on the momenta of the final leptons,
we calculate first the integral

\begin{equation}
\int d^{4}p_{1}\int d^{4}p_{2}\; \delta ^{(4)}\left(k'-p_{1}-p_{2}\right)\delta ^{+}\left(p_{1}^{2}-m^{2}\right)\delta ^{+}\left(p_{2}^{2}-m^{2}\right)\mathsf{L}_{\mu \rho },\label{eq:finalleptonmomentaintegral}\end{equation}
 where \emph{m} denotes the lepton mass and then contract it with
$H^{\mu \rho }$. The calculation of Eq.(\ref{eq:finalleptonmomentaintegral})
is particularly simple in the lepton pair center-of-mass frame. After
contraction, we still have to integrate over $d^{4}k'$ and $d^{3}p'$
and divide the expression by the flux factor. In terms of the Mandelstam
variables for the scattering process at the parton level, the subprocess
differential cross section reads

\begin{eqnarray}
\frac{d^{2}\sigma _{C}}{d\hat{t}dM^{2}} & = & -\left(\frac{2\alpha ^{3}Q_{a}^{4}}{3}\right)\frac{1}{\hat{s}^{2}}\left[\frac{\hat{s}^{2}+\hat{u}^{2}+2M^{2}\hat{t}}{\hat{s}\hat{u}}\, \right]\Phi \left(M^{2}\right),\label{eq:comptonsubprocesscrosssection}
\end{eqnarray}
 with $M^{2}$ being the invariant mass squared of the lepton pair
and $\Phi \left(M^{2}\right)=\frac{M^{2}+2m^{2}}{M^{4}}\sqrt{1-\frac{4m^{2}}{M^{2}}}\; \theta \left(M^{2}-4m^{2}\right).$
The cross section for photon-proton inelastic scattering in the parton
model is obtained by summing Eq.(\ref{eq:comptonsubprocesscrosssection})
over all types \emph{a} of partons and all possible longitudinal momentum
fractions \emph{x} weighted with the parton distribution functions
$f_{a}\left(x\right)$. Furthermore, we write $\hat{t}=-Q^{2}$ and
$\hat{s}=xs$ where \emph{s} represents the square of the photon-proton
center of mass energy. Since the mass of the scattered parton vanishes,
one gets $x=Q^{2}/\left[2P\cdot (k-k')\right]\equiv x_{B}$. The Compton
differential cross section for inclusive photoproduction of lepton
pairs can be then written as

\begin{eqnarray}
\frac{d^{3}\sigma _{C}\left[\gamma p\rightarrow l^{+}l^{-}X\right]}{dQ^{2}dM^{2}dx_{B}} & = & \frac{2\alpha ^{3}}{3}\int _{0}^{1}dx\; \delta \left(x-x_{B}\right)\; \sum _{a}f_{a}\left(x\right)Q_{a}^{4}\frac{1}{\left(xs\right)^{2}}\left[\frac{\left(xs\right)^{2}+\left(M^{2}+Q^{2}-xs\right)^{2}-2M^{2}Q^{2}}{xs\left(M^{2}+Q^{2}-xs\right)}\right]\Phi \left(M^{2}\right).\nonumber \\
 &  & \label{eq:comptonprocesscrosssection}
\end{eqnarray}

\section{The Bethe-Heitler process}

The Bethe-Heitler contribution can be calculated from Feynman diagrams
shown in Fig. \ref{diagrams}. The amplitude reads

\begin{eqnarray}
M_{BH}=M_{BH1}+M_{BH2} & = & -e^{3}Q_{a}\frac{1}{\left(k-k'\right)^{2}}u(p')\gamma _{\mu }u(p)\epsilon _{\nu }(k)\overline{u}(p_{1})\left[\frac{2p_{1}^{\nu }\gamma ^{\mu }-\gamma ^{\nu }\not \! k\gamma ^{\mu }}{-2\left(p_{1}\cdot k\right)}+\frac{\gamma ^{\mu }\not \! k\gamma ^{\nu }-2p_{2}^{\nu }\gamma ^{\mu }}{-2\left(p_{2}\cdot k\right)}\right]v(p_{2}),\nonumber \\
 &  & \label{eq:betheheitleramplitude}
\end{eqnarray}
 and moreover

\begin{eqnarray}
\frac{1}{4}\sum _{polarizations}|M_{BH}|^{2} & = & \frac{e^{6}Q_{a}^{2}}{4\left(k-k'\right)^{4}}H_{\mu \rho }L^{\mu \rho },\nonumber \\
H_{\mu \rho } & = & 4\left[p_{\mu }p'_{\rho }+p_{\rho }p'_{\mu }-g_{\mu \rho }\left(p\cdot p'\right)\right],\nonumber \\
L^{\mu \rho } & = & 4\left\{ \frac{A_{BH}}{\left(p_{1}\cdot k\right)^{2}}+\frac{B_{BH}}{\left(p_{2}\cdot k\right)^{2}}+\frac{C_{BH}}{\left(p_{1}\cdot k\right)\left(p_{2}\cdot k\right)}\right\} ,\nonumber \\
A_{BH} & = & \left(p_{1}\cdot k\right)\left[p_{2}^{\mu }k^{\rho }+p_{2}^{\rho }k^{\mu }-g^{\mu \rho }\left(p_{2}\cdot k\right)\right]-m^{2}\left[p_{1}^{\mu }p_{2}^{\rho }+p_{1}^{\rho }p_{2}^{\mu }-g^{\mu \rho }\left(p_{1}\cdot p_{2}\right)\right]\nonumber \\
 &  & +m^{2}\left[p_{2}^{\mu }k^{\rho }+p_{2}^{\rho }k^{\mu }-g^{\mu \rho }\left(p_{2}\cdot k\right)\right]-m^{2}\left(p_{1}\cdot k\right)g^{\mu \rho }+m^{4}g^{\mu \rho },\nonumber \\
B_{BH} & = & \left(p_{2}\cdot k\right)\left[p_{1}^{\mu }k^{\rho }+p_{1}^{\rho }k^{\mu }-g^{\mu \rho }\left(p_{1}\cdot k\right)\right]-m^{2}\left[p_{1}^{\mu }p_{2}^{\rho }+p_{1}^{\rho }p_{2}^{\mu }-g^{\mu \rho }\left(p_{1}\cdot p_{2}\right)\right]\nonumber \\
 &  & +m^{2}\left[p_{1}^{\mu }k^{\rho }+p_{1}^{\rho }k^{\mu }-g^{\mu \rho }\left(p_{1}\cdot k\right)\right]-m^{2}\left(p_{2}\cdot k\right)g^{\mu \rho }+m^{4}g^{\mu \rho },\nonumber \\
C_{BH} & = & 2\left(p_{1}\cdot p_{2}\right)\left[p_{1}^{\mu }p_{2}^{\rho }+p_{1}^{\rho }p_{2}^{\mu }-g^{\mu \rho }\left(p_{1}\cdot p_{2}\right)\right]-\left(p_{1}\cdot k\right)\left[p_{1}^{\mu }p_{2}^{\rho }+p_{1}^{\rho }p_{2}^{\mu }-2p_{2}^{\mu }p_{2}^{\rho }\right]\nonumber \\
 &  & -\left(p_{2}\cdot k\right)\left[p_{1}^{\mu }p_{2}^{\rho }+p_{1}^{\rho }p_{2}^{\mu }-2p_{1}^{\mu }p_{1}^{\rho }\right]-\left(p_{1}\cdot p_{2}\right)\left[p_{1}^{\mu }k^{\rho }+p_{1}^{\rho }k^{\mu }+p_{2}^{\mu }k^{\rho }+p_{2}^{\rho }k^{\mu }\right]\nonumber \\
 &  & +2g^{\mu \rho }\left(p_{1}\cdot p_{2}\right)\left[\left(p_{1}\cdot k\right)+\left(p_{2}\cdot k\right)-m^{2}\right]-2m^{2}k^{\mu }k^{\rho }.\label{eq:averagedbetheheitleramplitudessquared}
\end{eqnarray}
 Following the same procedure as in the previous section and after
some tedious algebra one obtains the subprocess differential cross
section,

\begin{eqnarray}
\frac{d^{2}\sigma _{BH}}{d\hat{t}dM^{2}} & = & \left(\alpha ^{3}Q_{a}^{2}\right)\frac{1}{\hat{s}^{2}\hat{t}^{2}\left(M^{2}-\hat{t}\, \right)}\left[-C_{1}\hat{u}\hat{s}+C_{2}\hat{t}\, \right].\label{eq:betheheitlersubprocesscrosssection}
\end{eqnarray}
 The coefficients in Eq.(\ref{eq:betheheitlersubprocesscrosssection})
are the functions of the invariants $m^{2}$, $M^{2}$ and $\left(k\cdot k'\right)=\left(M^{2}-\hat{t}\, \right)/2$
given by

\begin{eqnarray*}
C_{1} & = & \left[-4+\frac{6M^{2}}{\left(k\cdot k'\right)}-\frac{4M^{4}}{\left(k\cdot k'\right)^{2}}+\frac{M^{6}}{\left(k\cdot k'\right)^{3}}-\frac{20m^{2}M^{2}}{\left(k\cdot k'\right)^{2}}-\frac{4m^{4}M^{2}}{\left(k\cdot k'\right)^{3}}+\frac{6m^{2}M^{4}}{\left(k\cdot k'\right)^{3}}+\frac{16m^{2}}{\left(k\cdot k'\right)}+\frac{8m^{4}}{\left(k\cdot k'\right)^{2}}\right]\\
 &  & \times \ln \left[\frac{1-\sqrt{1-4m^{2}/M^{2}}}{1+\sqrt{1-4m^{2}/M^{2}}}\right],\\
C_{2} & = & \left[4\left(k\cdot k'\right)-4M^{2}+\frac{2M^{4}}{\left(k\cdot k'\right)}+\frac{4m^{2}M^{2}}{\left(k\cdot k'\right)}-\frac{8m^{4}}{\left(k\cdot k'\right)}\right]\ln \left[\frac{1-\sqrt{1-4m^{2}/M^{2}}}{1+\sqrt{1-4m^{2}/M^{2}}}\right]\\
 &  & +\left[4\left(k\cdot k'\right)-8M^{2}+\frac{4M^{4}}{\left(k\cdot k'\right)}+\frac{4m^{2}M^{2}}{\left(k\cdot k'\right)}\right]\sqrt{1-\frac{4m^{2}}{M^{2}}}.
\end{eqnarray*}
 Finally, one finds the Bethe-Heitler differential cross section for
inclusive photoproduction of lepton pairs namely,

\begin{eqnarray}
\frac{d^{3}\sigma _{BH}\left[\gamma p\rightarrow l^{+}l^{-}X\right]}{dQ^{2}dM^{2}dx_{B}} & = & \alpha ^{3}\int _{0}^{1}dx\; \delta \left(x-x_{B}\right)\; \sum _{a}f_{a}\left(x\right)Q_{a}^{2}\frac{1}{\left(xs\right)^{2}Q^{4}\left(M^{2}+Q^{2}\right)}\left[C_{1}\left(M^{2}+Q^{2}-xs\right)xs+C_{2}Q^{2}\right].\nonumber \\
 &  & \label{eq:betheheitlerprocesscrosssection}
\end{eqnarray}

\section{The interference terms}

Four Feynman diagrams give us eight interference terms. It turns out,
that they mutually cancel each other after being integrated over the
final lepton momenta. To prove this, let us focus only on the integrals
over the first two terms, i.e. $M_{C1}M_{BH1}^{\star }$ and $M_{C1}M_{BH2}^{\star }$
. They contain the following expressions

\begin{eqnarray}
\int d^{4}p_{1}\int d^{4}p_{2}\; \delta ^{\left(4\right)}\left(k'-p_{1}-p_{2}\right)\frac{1}{\left(p_{1}\cdot k\right)}\mathrm{Tr}\left[\left(\not \! p_{1}+m\right)\gamma _{\mu }\left(\not \! p_{2}-m\right)\gamma ^{\xi }\frac{\not \! p_{1}-\not \not \! k+m}{(p_{1}-k)^{2}-m^{2}}\gamma _{\nu }\right], &  & \nonumber \\
\int d^{4}p_{1}\int d^{4}p_{2}\; \delta ^{\left(4\right)}\left(k'-p_{1}-p_{2}\right)\frac{1}{\left(p_{2}\cdot k\right)}\mathrm{Tr}\left[\left(\not \! p_{1}+m\right)\gamma _{\mu }\left(\not \! p_{2}-m\right)\gamma _{\nu }\frac{\not \not \not \! k-\not \! p_{2}+m}{(k-p_{2})^{2}-m^{2}}\gamma ^{\xi }\right]. &  & \label{eq:firsttwointerferencetermsastraces}
\end{eqnarray}
 First, we use the delta function to integrate over $d^{4}p_{2}$.
After we perform the transposition of the trace and the momentum shift
$p_{1}-k'=-\tilde{p}$ of the second integrand in Eq.(\ref{eq:firsttwointerferencetermsastraces}),
the latter assumes the form

\begin{equation}
\int d^{4}\tilde{p}\; \frac{1}{\left(\tilde{p}\cdot k\right)}\mathrm{Tr}\left[\left(\not \! k'-\not \! \tilde{p}+m\right)^{\mathrm{T}}\gamma ^{\xi ^{\mathrm{T}}}\frac{\left(\not \not \not \! k-\not \not \! \tilde{p}+m\right)^{\mathrm{T}}}{(k-\tilde{p})^{2}-m^{2}}\gamma _{\nu }^{\mathrm{T}}\left(\not \! \tilde{p}-m\right)^{\mathrm{T}}\gamma _{\mu }^{\mathrm{T}}\right].\label{eq:integraloftraceovernewvariable1}\end{equation}
 Since the property $\hat{C}\gamma _{\mu }\hat{C}^{-1}=-\gamma _{\mu }^{\mathrm{T}}$
holds, where $\hat{C}$ represents the charge conjugation operator,
one ends up with

\begin{equation}
-\int d^{4}\tilde{p}\; \frac{1}{\left(\tilde{p}\cdot k\right)}\mathrm{Tr}\left[\gamma _{\mu }\left(\not \! k'-\not \! \tilde{p}-m\right)\gamma ^{\xi }\frac{\not \not \not \! \tilde{p}-\not \not \! k+m}{(k-\tilde{p})^{2}-m^{2}}\gamma _{\nu }\left(\not \not \! \tilde{p}+m\right)\right].\label{eq:integralovernewvariable2}\end{equation}
 The last expression is exactly equal but opposite in sign to the
first term in Eq.(\ref{eq:firsttwointerferencetermsastraces}). Thus,
by adding them we get zero. In fact, the result of cancellation is
known in general as the Furry's theorem: Feynman diagrams containing
a closed fermion loop with an odd number of photon vertices can be
omitted in the calculation of physical processes.

\section{Kinematics and Figures}

\begin{figure}[H]
\begin{center}

\includegraphics[  scale=1.2]{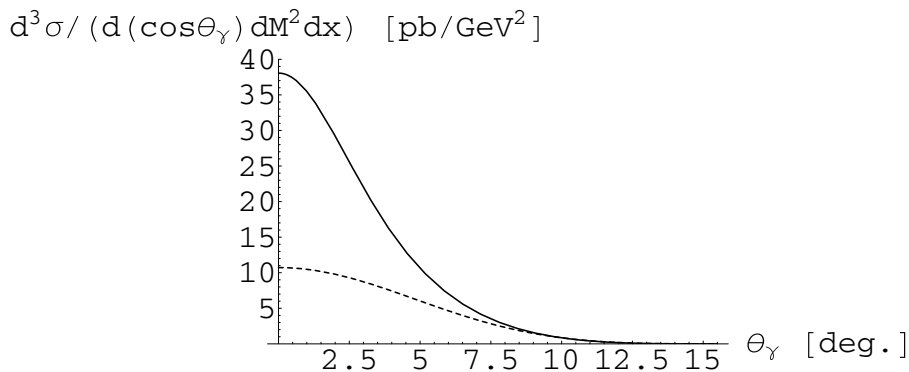}

\end{center}\caption{Differential cross section for Compton (solid line) and Bethe-Heitler (dashed line) contributions.}

\label{diffcrosssections}
\end{figure}

As $x\rightarrow 0$, the differential cross sections in Eqs.(\ref{eq:comptonprocesscrosssection})
and (\ref{eq:betheheitlerprocesscrosssection}) become singular. However,
since $s-M^{2}\geq \left(k-k'+P\right)^{2}\geq M_{p}^{2}$, one finds
in the laboratory frame, which is the rest frame of the proton with
mass $M_{p}$,

\begin{eqnarray}
\left[1+\left(2M_{p}E-M^{2}\right)/Q^{2}\right]^{-1} & \leq x\leq  & 1,\label{eq:rangeofx}
\end{eqnarray}
 where $E$ denotes the initial photon energy and $Q^{2}=2EE'-2E\sqrt{E'^{2}-M^{2}}\; \cos \vartheta _{\gamma }-M^{2}$
is the invariant momentum transfer. The energy of the pair and the
angle between photons are denoted by $E'$ and $\vartheta _{\gamma }$,
respectively. In Fig. \ref{diffcrosssections} both differential cross
sections are plotted against $\vartheta _{\gamma }$ at fixed values
of $E=40\, \mathrm{GeV}$, $M=3\, \mathrm{GeV}$ and $E'=10\, \mathrm{GeV}$.
Muons are taken as leptons and the following simplified parametrization
of parton distributions in the proton is used: $u_{val}\left(x\right)=1.89x^{-0.4}\left(1-x\right)^{3.5}\left(1+6x\right)$,
$d_{val}\left(x\right)=0.54x^{-0.6}\left(1-x\right)^{4.2}\left(1+8x\right)$
and $sea\left(x\right)=0.5x^{-0.75}\left(1-x\right)^{7}$ \cite{Radyushkin:1998rt}.

\begin{acknowledgments}
I would like to express sincere gratitude to my advisor Prof. A. Radyushkin
for his helpful comments. This work was supported by the US Department
of Energy DE-FG02-97ER41028 and by the contract DE-AC05-84ER40150
under which the Southeastern Universities Research Association (SURA)
operates the Thomas Jefferson Accelerator Facility.
\end{acknowledgments}


\begin{thebibliography}{1}
\bibitem{Brodsky:1972yx}S.~J.~Brodsky, J.~F.~Gunion and R.~L.~Jaffe, Phys.\ Rev.\ D
\textbf{6}, 2487 (1972). 
\bibitem{Radyushkin:1998rt}A.~V.~Radyushkin, Phys.\ Rev.\ D \textbf{58}, 114008 (1998) {[}arXiv:hep-ph/9803316{]}. \end{thebibliography}
\end{document}